\documentclass{aa}  

\usepackage{graphicx}
\usepackage{txfonts}
\usepackage{xcolor}
\usepackage{amsbsy,amsmath}
\usepackage{amsfonts}
\usepackage{natbib}
\usepackage[colorlinks,allcolors=blue]{hyperref}

\begin{document}

   \title{Planetary architectures under the influence of a stellar binary}

   \author{M. Ávila-Bravo
          \inst{1}\fnmsep\thanks{\email{mtavila@uc.cl}}
          \and
          C. Charalambous\inst{1}\fnmsep\thanks{\email{ccharalambous@uc.cl}}
          \and
          C. Aguilera-Gómez\inst{1}
          }

\institute{Instituto de Astrofísica, Pontificia Universidad Católica de Chile, Av. Vicuña Mackenna 4860, 782-0436 Macul, Santiago, Chile
        }

   \date{}

\titlerunning{Planetary architectures under the influence of a stellar binary}

 
  \abstract
{The presence of a stellar companion can strongly influence the architecture and long-term stability of planetary systems. Motivated by the discovery of exoplanets exhibiting extremely high eccentricities ($e \geq 0.8$) in systems with a binary companion, we investigate how planetary orbits around one star (S-type configuration) evolve under the gravitational perturbations of the companion.}
{We aim to assess the role of a stellar companion in shaping the orbital evolution of S-type planets and to explore whether dynamical interactions in such environments can account for the formation of highly eccentric planets.}
{We performed a suite of N-body simulations, modeling systems initially composed of three Jupiter-mass planets on nearly circular, coplanar orbits around the primary star. We systematically varied the semi-major axis, eccentricity, and inclination of the stellar companion, to characterize the conditions under which extreme eccentricities can be excited.}
   {Our results show that dynamical processes such as planet-planet scattering and secular mechanisms--including the von Zeipel-Kozai-Lidov effect induced by the binary--often act together to produce abrupt and significant changes in planetary orbital evolution, with the outcome strongly dependent on the binary separation. The binary's eccentricity primarily dictates the number of surviving planets, while its inclination not only governs the final eccentricities of those survivors but also drives their orbits to align with the binary plane. Our simulations successfully reproduce the high eccentricities and compact orbits observed in four observed systems, showing close agreement between the modeled configurations and the actual systems.}
   {}


   \maketitle
%
\section{Introduction}
\begin{figure*}
    \centering
    \includegraphics[width=0.47\textwidth]{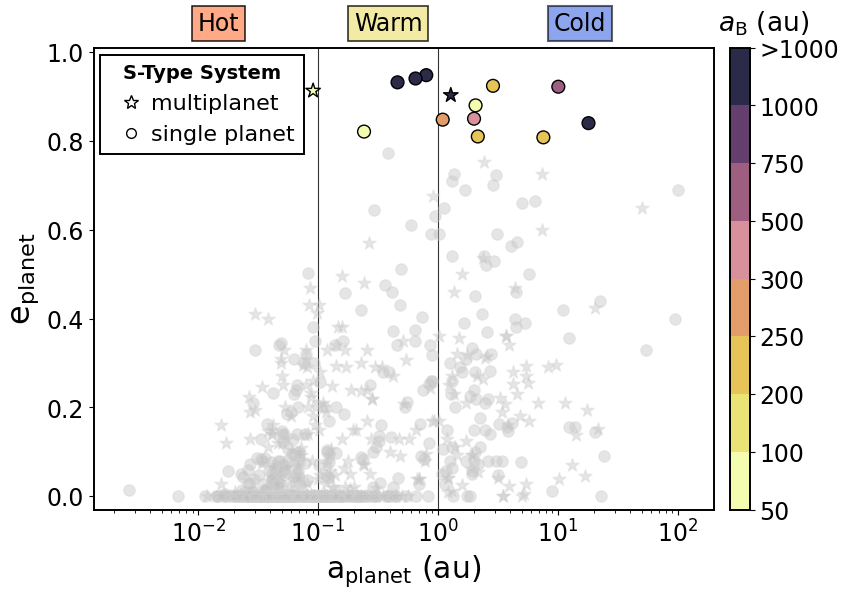}
    \quad
    \includegraphics[width=0.45\textwidth]{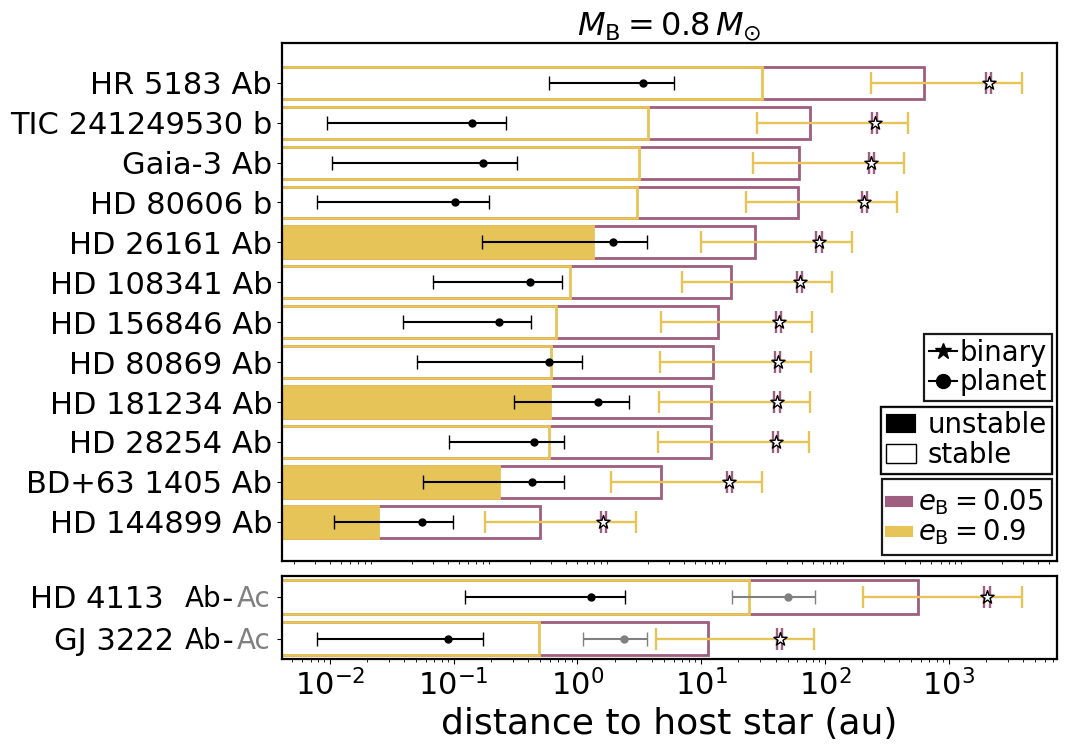}
    \caption{\textbf{Left:} Semi-major axis versus eccentricities of exoplanets in S-type orbits. Circles and stars represent systems with one or multiple planets, respectively. The color bar represents the semi-major axis of the binary star ($a_{\rm B}$) for planets with eccentricities greater than 0.8. Shaded regions highlight the hot, warm and cold Jupiter populations. Data extracted from the Encyclopedia of Exoplanets (\url{https://exoplanet.eu/catalog/}). \textbf{Right:} Orbital stability limits for systems with $e \geq 0.8$. The effects of a binary stellar companion with $M_{\rm B} = 0.8 \, M_{\odot}$ are shown for two eccentricities: $e_{\rm B} = 0.05$ (purple) and $e_{\rm B} = 0.9$ (yellow). The stars indicate the observed semi-major axis $a_{\rm B}$ of the binary, while the horizontal lines extend from periapsis $q_{\rm B} =a_{\rm B} (1 - e_{\rm B} )$ to apoapsis $Q_{\rm B} = a_{\rm B} (1 + e_{\rm B} )$. Similarly, the horizontal black circles and lines depict the semi-major axis, peri- and apoapsis of the planet, respectively. The purple and yellow bars extend to the critical semi-major axis $a_{\rm crit}$ for the planetary orbit, above which it would not be dynamically stable following \cite{Holman_1999}, for each binary eccentricity. Filled bars indicate that the system is dynamically unstable under the corresponding binary eccentricity, while open bars indicate stability.}
    \label{fig:catalogo_estabilidad}
\end{figure*}
It is now well established that multiple stellar systems are ubiquitous, particularly among young stars \citep{2023ASPC..534..275O}. Observational surveys indicate that about 65\% of stars are born not in isolation but as members of binary or higher-order multiple systems \citep{2013ARA&A..51..269D,2014prpl.conf..267R}, and roughly 40–45\% of Sun-like stars in the solar neighborhood remain in such systems at later stages \citep{2010ApJS..190....1R}. Among these, an estimated 10–15\% are known to host planets \citep[][]{2007A&A...462..345D, 2009A&A...494..373M,Roell_2012}. 

When studying planets in binary stellar systems, it is important to distinguish between circumstellar planets (S-type, orbiting one star while the other acts as a gravitational perturber) and circumbinary planets (P-type, orbiting both stars). In this work we focus exclusively on S-type configurations.

Stellar companions can alter both the formation and long-term evolution of planetary systems. Early on, they can truncate the circumstellar disk and induce gravitational perturbations \citep{1977ApJ...216..822P,10.1093/mnras/stv1450,Marzari_2019}, limiting where planets can form. Over longer timescales, several mechanisms were proposed to excite orbital eccentricities and induce planet migration, including the von Zeipel-Lidov-Kozai (vZLK) mechanism in stellar binaries \citep{Wu_2003,2007ApJ...669.1298F,2012ApJ...754L..36N,2015ApJ...799...27P}, planet-planet scattering \citep{1996Sci...274..954R,2011ApJ...742...72N,2012ApJ...751..119B}, and secular interactions between planets \citep{2011ApJ...742...72N,2011ApJ...735..109W}. 
The extent and outcome of these effects depend on the companion's separation, inclination, and eccentricity \citep{2008A&A...486..617K,Thebault_2015,2025Symm...17..344C}. Understanding the conditions under which a stellar companion alters planetary architectures is therefore essential for building a complete picture of planet formation and evolution in binary systems.

While theoretical studies show that planetary systems can remain stable around one component of a stellar binary despite periodic perturbations \citep[e.g.,][]{2016CeMDA.124..405A,Quarles_2020}, observations reveal a strong dependence of planet occurrence on binary separation $a_{\rm B}$. Exoplanets are found less frequently in `close binaries' \citep[separations $<100$\,au,][]{2007A&A...462..345D,2011IAUS..276..409E,10.1093/mnras/stab2328}, whereas `wide binaries' (with $a_{\rm B}>1500$\,au) show planet occurrence rates comparable to those around single stars \citep{Wang_2014b,Kraus_2016,2020Galax...8...16B}. However, the impact of stellar companions at intermediate separations (200-1500\,au) remains an open question \citep[e.g.,][]{Wang_2014b, 10.1093/mnras/stab2328, 2025arXiv250618759T} In this context, it is crucial to understand how the presence of a binary companion affects planetary system architectures and orbital stability. \citet{Thebault_2015} showed that, depending on the binary's orbital parameters, the characteristics of exoplanets in such systems can differ significantly from those formed around single stars.

Observational studies with TESS targets have suggested that binary companions can affect orbital alignment. \citet{Behmard2022} and \citet{Christian2025} found evidence for alignment between close-in planets and the stellar companion, with differences between small planets and gas giants. Hot Jupiters appear less aligned with the binary, although these results may be affected by false positives. Moreover, the binary separation plays a critical role, as \citet{Christian2022} found alignment for binaries with separations $<700$ au, but not at larger separations.

Altogether, this growing observational evidence highlights the impact of binarity on both the occurrence and the architecture of planetary systems. Yet, most studies are intrinsically constrained, as they typically focus on close-in planets and on identifying stellar companions to known planet hosts, rather than surveying planetary systems around binaries selected independently of planetary presence. In a complementary approach, \citet{Su2021} analyzed planets detected through radial velocity and showed that multi-planet systems tend to occur around wider binaries. Their results suggest that planet formation and evolution in binaries with separations $>100-300$ au may resemble that of single stars.

The left panel of Fig.~\ref{fig:catalogo_estabilidad} shows observed S-type planets (data from the \texttt{Encyclopedia of exoplanetary systems} \citep[\url{https://exoplanet.eu/planets_binary}, ][]{Schneider_2011,2025arXiv250618759T} as of August 2025). In the planetary semi-major axis-eccentricity plane, we highlight a small group of highly eccentric planets ($e_p\geq0. 8$). These systems challenge standard planet formation theories, as planets forming in protoplanetary disks are expected to have low eccentricities due to dissipative processes that tend to circularize their orbits \citep{Ford_2001}. Moreover, statistical studies restricted to single stars show that in single-planet systems have an average eccentricity of $\sim$0.3 \citep{2008ApJ...685..553S}, whereas multi-planet systems tend to have nearly circular orbits, with average eccentricities of about 0.05 \citep{2019AJ....157...61V}. In wide binaries, planetary eccentricities tend to be larger than those observed around single stars, with no significant difference between single- and multiple-planet systems. This effect remains detectable even in very wide binaries with separations exceeding $\sim1000$\,au \citep{Su2021}. This points to a key role of dynamical interactions with an outer perturber in the generation of extreme planet eccentricities \citep{Mustill_2022}.

Observations of systems such as \texttt{HD~80606}, \texttt{HD~80869} or \texttt{HD~28254}—each containing a Jupiter-mass planet with eccentricity $e>0.8$ and a wide stellar companion—suggest that dynamical interactions with outer perturbers play a key role in generating extreme planetary eccentricities \citep{2001A&A...375L..27N,  Desidera_2006, 2008A&A...480L..33T, 2018A&A...614A..16C, Mustill_2022}.
These effects are characteristic of hierarchical triple systems, where a planet of mass $m_p$ orbits a primary star of mass $M_\star$, that is itself orbited by a distant stellar companion of mass $M_{\rm B}$, with $M_\star \sim M_{\rm B} \gg m_p$. In such configurations, a sufficiently inclined binary orbit can induce the von Zeipel-Lidov-Kozai  mechanism \citep{1910AN....183..345V,1962P&SS....9..719L,1962AJ.....67..591K}, which drives long-term secular oscillations in eccentricity and inclination while leaving the semi-major axis nearly unchanged. This dynamical process, absent in isolated star-planet systems, can dramatically reshape planetary orbits over long timescales, potentially producing close-in, highly eccentric, or misaligned hot Jupiters.

Observational data typically provide the projected separations between the stars, while the rest of the orbital parameters are largely unconstrained, mainly due to the long orbital periods. However, using Gaia DR3, \cite{Hwang_2022}
found that close binaries ($a_{\rm B}<100$\,au) have uniform eccentricity distribution, while wide binaries ($a_{\rm B}>1000$\,au) are highly eccentric. In the right panel of Fig.~\ref{fig:catalogo_estabilidad}, we illustrate the stability limits following \citet{Holman_1999}, by considering two extreme cases for the binary eccentricity and assuming a companion mass of $M_{\rm B} = 0.8 \, M_\odot$, for the 14 systems with $e_p>0.8$.\footnote{The sample includes 14 systems, two of which host multiple planets. It is important to note that the critical semi-major axis formula from \citet{Holman_1999} was derived under the assumption of a single-planet system. Therefore, its application to multiplanet systems should be interpreted with caution, as mutual gravitational interactions between planets could either enhance or reduce the overall stability depending on the specific architecture.}
If the companion has a low eccentricity ($e_{\rm B} =0.05$, purple), all known single-planet systems lie well within the stable region (empty boxes), suggesting they can remain dynamically stable over long timescales under the influence of a moderately circular binary. However, in the case of a highly eccentric binary ($e_{\rm B} =0.9$, shown in yellow), several planets would lie beyond the critical semi-major axis (filled boxes), placing them in a regime where the gravitational perturbations from the stellar companion could destabilize their orbits on relatively short timescales. This motivates our exploration of how the long-term secular effects of binary companions in different orbital configurations may contribute to the dynamical excitation of planets in otherwise marginally stable systems. 

To investigate the effect of binaries in different configurations, we simulate systems with three planets subject to mutual gravitational interactions and the influence of a distant stellar companion. Planet-planet scattering, a key mechanism shaping planetary systems after formation \citep{1996Natur.384..619W}, can be triggered even by mild perturbations \citep[such as those generated by stellar flybys,][]{2025A&A...696A.175C,2025arXiv250808406W} and is known to excite eccentricities. However, this mechanism alone is insufficient to explain the observed high eccentricities. 

Motivated by the small but intriguing population of highly eccentric S-type planets, we investigate whether such configurations can naturally arise from the combined effects of early planet-planet scattering and long-term secular perturbations from a stellar companion. Our study builds upon previous works \citep{Marzari_2021,stegmann2025,2025ApJ...980L..31W} that examined the dynamical consequences of binary-induced secular evolution.

We perform a suite of N-body simulations, initializing three giant planets on nearly circular orbits, consistent with formation in a truncated circumstellar disc. After an initial phase of planet-planet scattering on $\sim$Myr timescales, the systems evolve under the gravitational influence of a distant stellar companion. This companion can trigger vZLK oscillations which transfer angular momentum between inclination and eccentricity, potentially pushing planets into the highly eccentric regime observed in S-type systems over $\sim$Myr-Gyr timescales. Our approach combines both scattering and secular forcing in a self-consistent framework, enabling us to probe the outer regions of the eccentricity distribution of Warm and Cold Jupiters.

Rather than revisiting the well-studied origins of Hot Jupiters \citep[e.g.,][]{1996Sci...274..954R,Zhou_2007,2008ApJ...678..498N,2025ApJ...980L..31W}, our focus lies in identifying the binary configurations capable of producing the high eccentricities observed in longer-period gas giants. Our study is particularly motivated by the relatively unexplored regime of intermediate binary separations (200–1000 au), where the dynamical interplay between scattering and secular effects remains poorly understood. In Section~\ref{sect:configuracion}, we describe our simulation setup and initial conditions. Section~\ref{sect:general_results} presents the results of our general parameter exploration, while in Section~\ref{sect:aplication_observed_systemas} we apply our framework to observed systems with known binary companions. We conclude with a discussion  of our findings and their broader implications in Section \ref{sect:discussion}, and briefly summarize our results in Section~\ref{sect:conclusiones}.

\section{Setup of the simulations}\label{sect:configuracion}

In this section we outline the setup and initial conditions for our simulations. Following the approach in \cite{Marzari_2021}, 
and motivated by the theoretical expectation that protoplanetary disks rarely form only a single body \citep{2010apf..book.....A}, 
we initialize a system consisting of three Jupiter-mass planets in S-type orbits around a central star of mass $M_\star = 1\,M_\odot$. An external stellar companion with mass $M_{\rm B}= 0.8 \,M_\odot$ is included, with the primary star taken as the origin of the reference frame. To track the long-term evolution of the system, we perform N-body simulations using the \texttt{REBOUND} code \citep{Rein_2012} with the IAS15 integrator. 

To explore how gravitational perturbations induced by a stellar companion affect the long-term architecture of planetary systems, we adopt a physically motivated setup in which the planets are initially placed on dynamically cold orbits: low-eccentricities ($e_i \leq 0.01$, being $i=1,2,3$) and moderate inclinations ($i_i \leq 3^\circ$), consistent with expectations from formation in a protoplanetary disk \citep[see][and references therein]{2010ASSL..366..135K}. The planets are  placed at semi-major axes $(a_1,a_2,a_3) = (3, 6, 9)$ au. Our primary goal is to determine whether perturbations from the binary companion can trigger dynamical instabilities, potentially leading to collisions or ejections, and ultimately resulting in single-planet systems on highly eccentric orbits--configurations that resemble certain observed exoplanetary systems (see Fig.~\ref{fig:catalogo_estabilidad}).

This setup is motivated by both theoretical considerations and observational trends. Systems with three giant planets could plausibly form even in the presence of a moderately separated binary companion \citep{Marzari_2021}. Additionally, the planetary population observed in binaries is dominated by Jupiter-mass planets \citep{Quarles_2020}, supporting our choice of equal-mass giants. Finally, the scarcity of wide-orbit planets \citep[$a_{p} > 10$ au,][]{Martin_2018} in current surveys may reflect not only observational biases--particularly of radial velocity and transit techniques--but also dynamical sculpting by binary companions, which we aim to explore through our simulations.

Collisions are handled using the \texttt{sim.collision\_resolve} method implemented in \texttt{REBOUND}\footnote{\url{https://rebound.readthedocs.io/en/latest/collisions/}}, which monitors the distance between particles at each timestep. A collision is considered to occur when the distance $d$ between two planets is less than the sum of their radii: $d < r_i + r_j$. Each planet’s radius is assigned to 1 Jupiter radius.
When a collision is detected, the two planets merge into a single object with mass $m_i + m_j$, and the orbit of the resulting planet is then calculated using conservation of momentum. 

We consider a planet to be ejected if it simultaneously satisfies two conditions: (i) its distance from the central star exceeds 50 au, and (ii) its orbital eccentricity $e_i \geq 1$.

As mentioned above, the binary system consists of a solar-type primary and a slightly less massive companion ($0.8\ \mathrm{M_\odot}$). This configuration reflects our intent to model a system with comparable but non-identical stellar masses. Observational studies show that binaries often exhibit high mass ratios, with an excess of nearly equal-mass systems \citep[$q=M_2/M_1\gtrsim0.95$,][]{El_Badry_2019, Hwang_2022}. However, the magnitude of this excess depends on the primary mass and tends to be more pronounced for lower-mass primaries.
Moreover, giant planets are more frequently found around G-type stars \citep[e.g.,][]{2005PThPS.158...68I, Mulders2018, Pan2025} making a $1.0\ \mathrm{M_\odot}$ primary astrophysically motivated for our purposes. Consequently, we adopted a slightly less massive companion, consistent with the observed trends for systems in this mass regime and separations of a few hundred au. We note that the adopted companion mass is a conservative choice, as a more massive companion is expected to enhance the dynamical effects discussed in this work.

We thus consider ten values for the binary semi-major axis, $a_{\rm B}$, ranging from 50 to 1500 au. This range is selected based on observational trends in planetary systems hosted by binary stars. Multiple studies have shown that the occurrence rate of planets is significantly lower in close binaries \citep[i.e., $a_{\rm B} < 50$ au,][]{Martin_2018}, with less than 10\% of known exoplanets found in such systems \citep{Thebault_2015}. In contrast, most planet-hosting binaries exhibit wider separations, typically exceeding 500 au \citep{Roell_2012}. At very large separations ($a_{B} \gtrsim 1500$ au), a modest decline in planet occurrence has also been reported \citep{Martin_2018}, suggesting this value as a practical upper limit for our parameter space.

It is important to simulate both close and wide binary systems, as the orbital eccentricity $e_{\rm B}$ distribution varies significantly with separation. According to \cite{Hwang_2022}, binaries with $a_{\rm B} < 100$ au exhibit an approximately uniform eccentricity distribution, showing no strong preference for low or high eccentricities. However, as separation increases, the distribution gradually shifts toward a superthermal regime, where high eccentricities become increasingly common. For $a_{\rm B} > 1000$ au, wide binaries exhibit highly elongated orbits.

In addition to separation and eccentricity, we also explored five configurations for the binary inclination, $i_{\rm B}$, ranging from fully coplanar to highly inclined systems. By varying these parameters, our aim is to identify the combinations that best reproduce the dynamical features of the observed planetary systems highlighted in Fig.~\ref{fig:catalogo_estabilidad}. We do not consider $i_{\rm B} > 81^\circ$, since such extreme inclinations are theoretically associated with the formation of hot Jupiters, whereas our focus here is on the population of cold Jupiters.

\begin{figure*}
\centering
    \includegraphics[width=\linewidth]{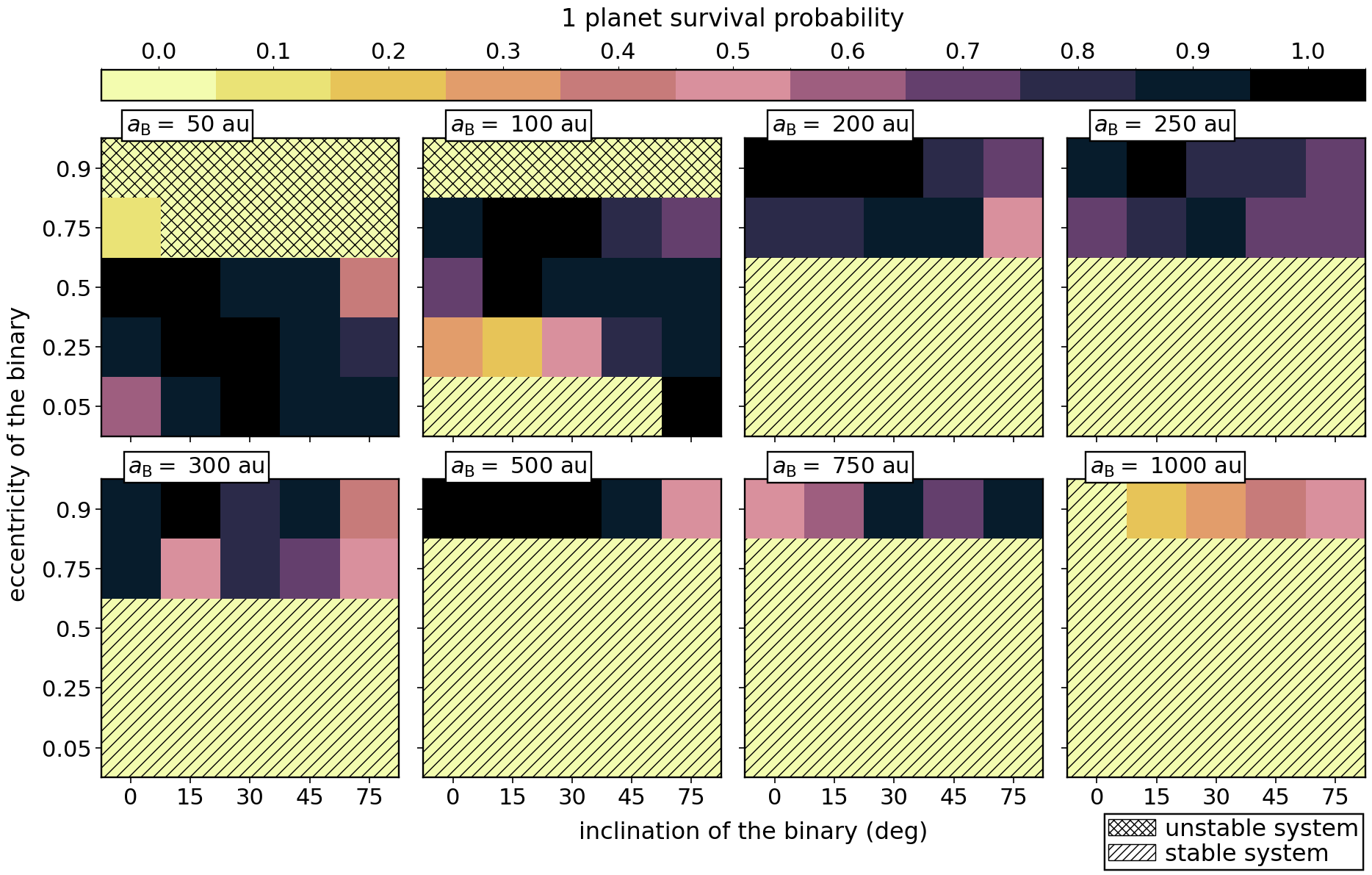}
    \caption{Survival probability of a single planet in a binary system that initially hosted three planets, across various orbital configurations. Each panel corresponds to a specific value of the binary's semi-major axis $a_{\rm B}$, with the x- and y-axes showing the binary's inclination $i_{\rm B}$ and eccentricity $e_{\rm B}$, respectively. Shaded regions with crosses denote dynamically unstable configurations where no planets survive, while diagonal hatches indicate stable configurations in which all three initial planets remain bound throughout the simulation.  }
    \label{probabilidad_1}
\end{figure*}

Our simulations are designed to systematically explore the impact of a binary companion on the stability of planetary systems. While the planetary orbits remain fixed in all runs, we vary the binary’s orbital parameters:  the semi-major axis, eccentricity and inclination. Specifically, we consider ten values of binary separation: $a_{\rm B} \in [50, 100, 200, 250, 300, 500, 750, 1000, 1250, 1500]$ au, five values of eccentricity: $e_{\rm B} \in [0.05, 0.25, 0.5, 0.75, 0.9]$, and consider 5 possible orbital inclinations relative to the initial midplane of the star-planet system: $i_{\rm B} \in [0^{\circ}, 15^{\circ}, 30^{\circ}, 45^{\circ}, 75^{\circ}]$. 

For each combination of $a_{\rm B}$, $e_{\rm B}$ and $i_{\rm B}$, we perform 10 simulations, each integrated up to $5\times10^{8}$ years. 
In each integration, the argument of periapsis $\omega_i$ of the planets is randomly chosen between 0 and 2$\pi$, allowing for a range of initial orbital geometries. This approach ensures that our results are not biased by any particular configuration but instead represent the general dynamical behavior of the system. In total, this yields 2500 simulations.

\section{Results}
In this section, we investigate orbital stability across the range of binary configurations described in Section \ref{sect:configuracion}.  Section~\ref{sect:general_results} presents a general mapping of the parameter space, while Section~\ref{sect:aplication_observed_systemas} applies these findings to specific observed systems.

\subsection{Parameter exploration}\label{sect:general_results}
\begin{figure*}
\centering
    \includegraphics[width=\linewidth]{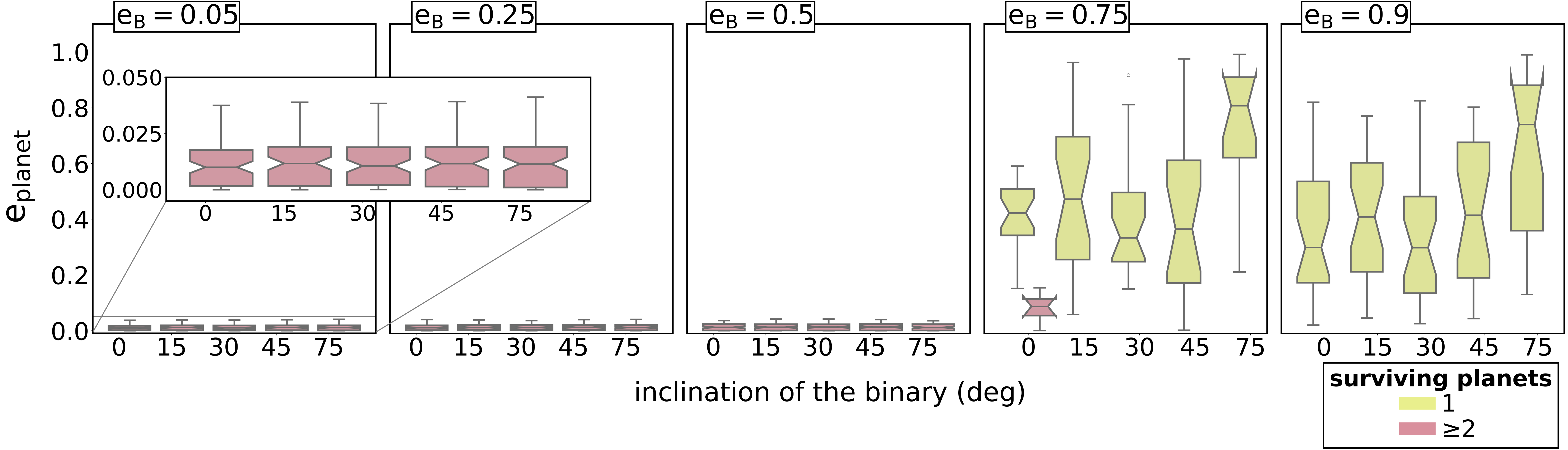}
    \caption{
    Planetary eccentricities as a function of binary inclinations for a fixed separation, $a_{\rm B} = 200$\,au. Each panel correspond to a specific value of $e_{\rm B}$. Purple boxes indicate the maximum eccentricities when more than two planets survive, while the yellow boxes show the eccentricities of systems with a single surviving planet.
    } 
    \label{box_plot}
\end{figure*}

The orbital parameters of the binary companions are largely unconstrained. Observational data typically provide only the projected separation, without direct information on the full set of orbital elements. However, as analytically suggested by \citet{1968BAN....20...47V} and more recently confirmed through Monte Carlo simulations by \citet{2025A&A...698A.102R}, the projected separation can be considered a reasonable proxy for the binary’s semi-major axis. Beyond that, the remaining orbital parameters—such as eccentricity and inclination, remain unknown. Additionally, the mass of the binary companion is often uncertain as well. 
Given these observational limitations, we adopt the numerical approach described in Section~\ref{sect:configuracion}, that allows us to isolate the effect of the companion star on planetary stability by modifying only the binary parameters.

The results of this parameter exploration are summarized in Fig.~\ref{probabilidad_1}, which shows the survival probability of a single planet over a grid of binary eccentricities and inclinations. Each panel corresponds to a different fixed binary separation. For each combination of $e_{\rm B}$ and $i_{\rm B}$, we run a set of 10 simulations. The color in each cell indicates the fraction of runs (out of 10) that result in the survival of exactly one planet following the gravitational interaction with the companion star. Overlaid hatching highlights globally unstable configurations—where all three planets are lost (cross-hatching)—and globally stable cases—where the three planets survive (diagonal-hatching)—following the classification described above. 

Our analysis is primarily qualitative, aiming to illustrate how different parameters of the binary companion affect planetary evolution in different ways, rather than to determine precise quantitative stability limits. For this reason, the parameter grid was designed to span representative values that capture the main dynamical trends of the system, rather than to exhaustively explore a denser grid of parameters.

At very close separations, systems are generally less stable. In the case of  $a_{\rm B}=50$\,au, for eccentricities $e_{\rm B}\geq0.75$ no planets survive regardless of the inclination, as indicated by the cross-hatching. For lower eccentricities ($e_{\rm B}\leq0.5$), survival is typically limited to a single planet with probabilities ranging from 0.8 to 1. This extends the results of \citet{Marzari_2005}, who also identified close binaries as a driver of planet-planet scattering leading to single eccentric giants.

For slightly bigger separations, when $a_{\rm B}=100$\,au, there is a slight improvement in stability compared to $a_{\rm B}=50$\,au, but unstable regions (cross-hatching) still dominate for high eccentricities. However, in this case, a nearly circular binary companion allow for all three planets to remain stable for $i_{\rm B}\leq 45^\circ$.
\begin{figure*}
\centering
    \includegraphics[width=1.0\textwidth]{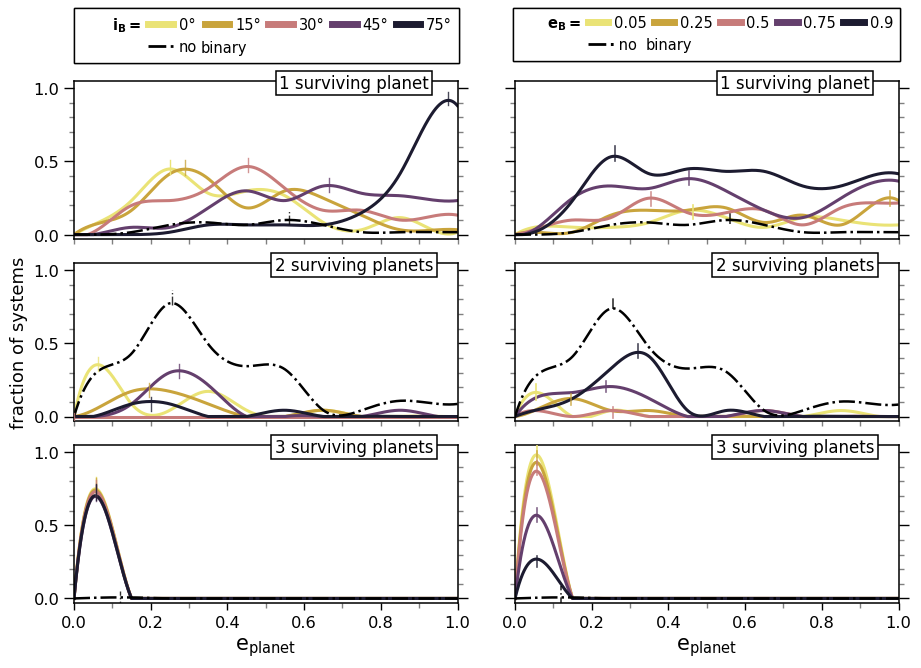} 
    \caption{Normalized distribution of planetary eccentricities from simulations of systems with varying binary inclinations (left) and binary eccentricities (right). Each row shows results for different survival outcomes: 1, 2, or 3 remaining planets. Colored curves correspond to individual values of $i_{\rm B}$ or $e_{\rm B}$, while the black dash-dotted line indicates the reference case without a binary companion. 
    }
    \label{fig:distribucion_inc_ecc}
\end{figure*}

This is consistent with the analytical stability criterion of \citet{Holman_1999}, which, although developed for systems with a single planet, sets a clear boundary between stable and unstable configurations. In our simulations, the planets in these unstable systems lie well beyond the predicted critical semi-major axis $a_{\rm crit}$, making rapid instability inevitable. On the other hand, at larger binary separations, $a_{\rm crit}$ increases markedly, and the planets start well within the stability limit, resulting in largely stable configurations. 

In contrast, for intermediate separations ($200\leq a_{\rm B}\leq300$\,au), only a highly eccentric binary ($\gtrsim 0.75$) lead to substantial perturbations. In the case of wide binaries ($a_{\rm B}\geq500$\,au), the effects become increasingly subtle, requiring very high eccentricities ($\gtrsim 0.9$) to noticeably affect planetary stability—though even then, the impact remains relatively modest. In our simulations, companions at 1250 and 1500\,au do not significantly perturb the system. However, \citet{2013Natur.493..381K} showed that at such separations the galactic tide \citep{1996MNRAS.282.1064H} can secularly alter the binary orbit, leading during close pericentre passages to the excitation of planetary eccentricities or even ejections. This highlights that, at these distances, considering only the direct gravitational interactions between the stars may be insufficient, and the effect of the galactic tide must also be taken into account. We also note that, as shown in \cite{2013Natur.493..381K} for $a_{\rm B}\geq1000$\,au, if the planets were initially located farther out, the outcome could differ, with more extended planetary systems eventually yielding much more excited eccentricities compared to compact ones.

As the binary separation increases, the parameter space allowing for the system to remain long-term stable broadens significantly. 
Specifically, for $a_{\rm B}=200$\,au and $e_{\rm B}\leq0.5$, stable outcomes with three surviving planets become common across all tested inclinations, maintaining the almost circular configurations until the end of the integration (see the first three panels in Fig.~\ref{box_plot}). 

In summary, Fig.~\ref{probabilidad_1} shows that although the influence of the binary diminishes as its separation increases, distant companions at $a_{\rm B}=1000$\,au can still affect the long-term orbital stability of the system, even without considering galactic tides. However, the conditions required for this influence change with distance. For close binaries, nearly any eccentricity and inclination can significantly destabilize the system. In particular, for small separations and high eccentricities, the system becomes completely unstable, with no planets surviving the interaction. 

Note that our results differ from those of \citet{Malmberg_2007}, who fixed the binary eccentricity at a moderate value for $a_{\rm B}=300$\,au, suppressing secular evolution in coplanar cases. In their simulations, planetary systems resembling the Solar System typically became unstable when the binary inclination exceeded $\sim 40^\circ$, triggering Kozai cycles that drove the outer planet to high eccentricity and subsequent planet-planet scattering, often leading to ejections. However, this outcome cannot be generalized. In our runs, using the same parameters and a more massive companion, all planets survived. Moreover, varying the binary eccentricity drastically alters the system's evolution, producing strong secular effects even in nearly coplanar systems. Overall, binary eccentricity mainly controls planet retention, while binary inclination shapes the eccentricities of the survivors.

Fig.~\ref{box_plot} provides additional insight into the dynamical outcomes by displaying the distribution of final planetary eccentricities as a function of binary inclination, for each of the tested binary eccentricities. This figure focuses on simulations with $a_{\rm B}=200$\,au, a representative case of intermediate separation. We distinguish between systems in which only one planet survives (shown in yellow) and those where two or more planets remain bound (in violet).

This plot clearly illustrates a dichotomy in the resulting planetary eccentricities: when multiple planets survive, they tend to retain low eccentricities ($e_p\lesssim0.05$), showing no signs of transient dynamical instability, regardless of the binary inclination. For this to happen, it is necessary that the binary eccentricity is low-to-moderate. In contrast, systems where only a single planet survives exhibit much higher final eccentricities, particularly in scenarios involving high binary inclination ($i_{\rm B}\geq 30^\circ$). This trend reflects the violent dynamical evolution leading to the loss of planets and suggests that high eccentricities in surviving planets may serve as a signature of past strong interactions with a perturber. The increase in planetary eccentricity becomes most pronounced at bigger binary eccentricity, especially at moderate-to-high inclinations. 

Complementing the parameter exploration presented in Fig.~\ref{probabilidad_1} and Fig.~\ref{box_plot}, Fig.~\ref{fig:distribucion_inc_ecc} shows the final eccentricity distribution of the surviving planets, separated by the number of remaining planets (1--3, from top to bottom) across all binary separations, considering variations in both binary inclination (left panels) and binary eccentricity (right panels). For comparison, we also include a control set of simulations without a binary companion, shown in the same plots (dashed dotted black lines). In these cases, the planets were initialized with slightly higher eccentricities, randomly drawn from the range $0.01 < e_p < 0.3$, to test whether planet-planet interactions alone could drive significant eccentricity growth. 

A robust result is that systems retaining all three planets remain dynamically cold, with eccentricity distributions sharply peaked at nearly zero values ($e_p \lesssim0.05$) across the full range of $i_{\rm B}$ and $e_{\rm B}$. This suggests that the presence of multiple planets in an observed system likely implies a relatively quiescent dynamical history, with little perturbation from the stellar companion. The differing curve heights reflect variations in the frequency of survival for each $e_{\rm B}$, rather than significant changes in the eccentricity distribution. In fact, mutual planet-planet interactions can induce rapid apsidal precession, which effectively suppresses vZLK cycles that would otherwise excite eccentricities \citep[see][]{Takeda_2008,2011A&A...533A...7B, 2022A&A...664A..87R}.

When interactions with the binary lead to the loss of one or two planets, the eccentricity distribution changes drastically. High binary inclinations tend to produce the most eccentric planets when only a single planet survives. Large binary eccentricities also excite planetary orbits, although the effect is milder. Systems with multiple survivors typically retain low eccentricities regardless of $i_{\rm B}$ or $e_{\rm B}$.

In systems where only a single planet survives, the evolution is far more violent: the distributions develop extended tails up to $e_p\simeq 1$, with an increasing fraction of highly eccentric planets as $i_{\rm B}$ or $e_{\rm B}$ grows. In particular, large inclinations ($\gtrsim45^\circ$) are the most efficient at producing highly eccentric surviving planets, consistent with planet-planet scattering and vZLK interactions triggered by strong stellar perturbations.
Systems with two surviving planets occupy an intermediate regime: their eccentricity distribution is broader than in the fully stable case but still peaks at lower values, suggesting moderate dynamical excitation.

\begin{figure}
\centering
    \includegraphics[width=\linewidth]{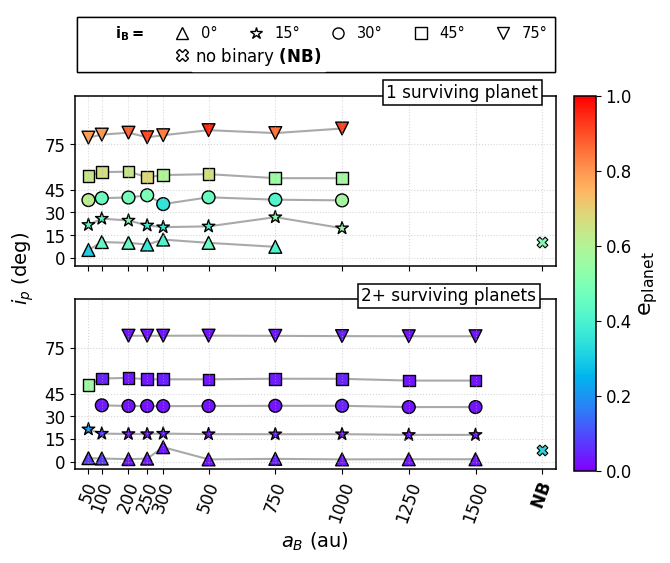}
    \caption{Mean planetary inclinations. 
    For each $a_{\rm B}$, all $e_{\rm B}$ are considered. The control case without a binary (NB) is shown as a cross. Each panel corresponds to a different survival outcome: systems ending with a single planet (top), or with two or three planets remaining (bottom). Symbols denote the initial inclination, while the colored bar indicates the averaged attained planetary eccentricity.} 
    \label{distribucion_incs}
\end{figure}

For isolated systems, \cite{Zhou_2007,2008ApJ...686..603J} suggest that such eccentricities may arise from scattering between multiple planets, while \cite{Carrera_2019} argue that this process could account for all observed eccentric planets, implying that vZLK excitation is not strictly required. However, in our simulations without a stellar companion, systems show a markedly different behavior: $e_p$ remains low in all cases. Thus, our results point in the opposite direction: stellar perturbations, rather than scattering, emerge as the dominant driver of eccentricity excitation, since planet-planet interactions alone are insufficient to reproduce the high-eccentricity tails observed in binaries. This highlights the binary's critical role in shaping planetary architectures.

Fig.~\ref{distribucion_incs} shows the average orbital inclination of the surviving planets relative to their initial plane ($i_{p}$), together with their maximum eccentricity, separated by the number of surviving planets and by the initial binary separation. In agreement with \citet{Zhang_2018}, planets generally tend to align with the binary's orbital plane. When multiple planets survive, the system remains well-ordered and eccentricities stay low. In contrast, if only a single planet survives, inclinations are slightly higher than the binary plane and show large oscillation--consistent with vZLK dynamics--while eccentricities increase with the binary's initial inclination. Simulations without a binary (NB) shown for comparison, retain their low eccentricities and stay close to their initial plane. Our results further indicate that highly inclined stellar companions play a decisive role in producing a population of highly eccentric single planets, challenging the interpretation based solely on single-star systems.

This broader view reinforces the previously discussed trend: single-planet systems with high eccentricities can be signatures of violent past dynamical evolution driven by stellar companions, while compact, nearly circular multiplanet systems trace dynamically calmer environments. Overall, this trend is more clearly associated with variations in binary inclination, while the effect of binary eccentricity is comparatively weaker and produces outcomes more akin to the control simulations without a binary companion.
\begin{figure*}
    \centering
    \includegraphics[width=\linewidth]{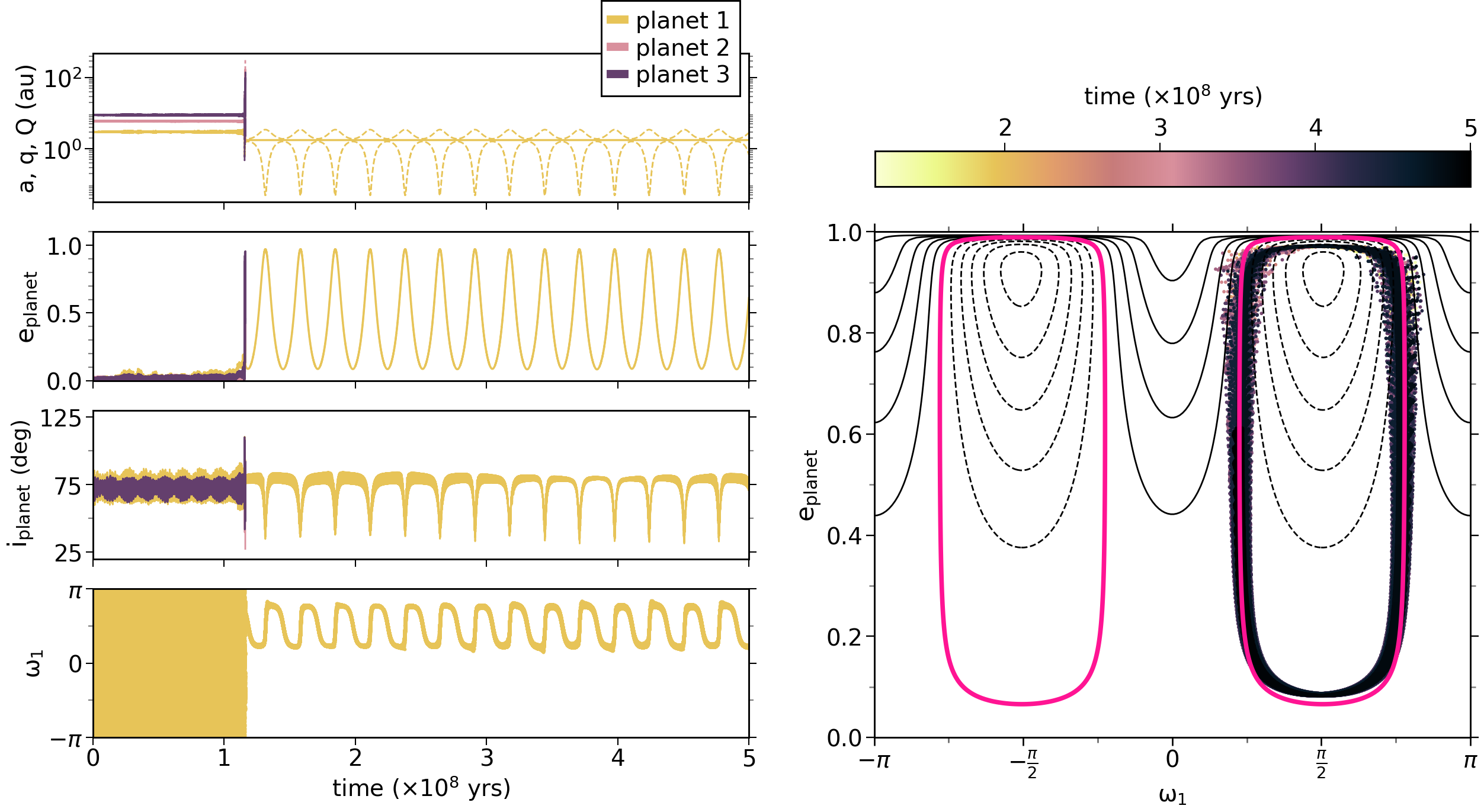}
    \caption{Temporal evolution of the simulation with $ a_{\rm B} = 1000$\,ua, $ e_{\rm B} = 0.9$, and $ i_{\rm B} = 75^{\circ}$. The left panels show, from top to bottom, the orbital elements of the planets: distance to the central star, $e_{j}$, and $i_j$, with $j = 1, 2, 3$. The bottom panel shows the angle associated to the vZLK resonance ($ \omega_{j} = \varpi_{j} - \Omega_{j}$). The right panel corresponds to the vZLK representative plane ($\omega_{j},e_{j}$), where the level curves of the octupolar Hamiltonian are superimposed with the simulation color-coded by time from the moment where the system is left with one planet. }
    \label{fig:resonance_comparison}
\end{figure*}

Fig.~\ref{fig:resonance_comparison} illustrates the temporal evolution of a representative simulation in which the binary companion is placed at a distance of 1000\,au, with high eccentricity ($e_{\rm B}=0.9$) and inclination ($i_{\rm B} = 75^\circ$). The left panels show the orbital elements. 
From top to bottom we display the semi-major axis along with the peri- ($q_i=a_i(1-e_i)$), and apocenters ($Q_i=a_i(1+e_i)$) of the planets, followed by the eccentricities, and inclinations as a function of time. The bottom panel shows the evolution of $\omega_1 = \varpi_1 - \Omega_1$, with $\varpi_1$ the argument of pericenter and $\Omega_1$ the longitude of the ascending node of planet 1. 
All quantities are measured with respect to the invariable plane of the system (also referred to as the Laplace plane), which in our configuration nearly coincides with the plane of the stellar binary, as it is perpendicular to the total angular momentum of the system \citep{2009MNRAS.400.1373L}.

Initially, the system contains three planets and remains stable for the first $\sim 10^8$\,yr. 
A brief phase of planet-planet scattering then drives the ejection of the two outer planets. 
This instability arises from long-term mutual interactions, amplified by perturbations from the binary companion. Control simulations without the stellar companion, using the same initial configuration, confirm that the system is stable at $t=0$, since the mutual separations ($\sim$7 mutual Hill radii) ensure short-term stability \citep{1996Icar..119..261C,2008ApJ...686..580C}. Over longer timescales, planet–planet interactions alone can generate dispersion; however, the resulting architectures do not reproduce the highly eccentric planets observed. As eccentricities increase, close encounters ultimately remove the outer two bodies, leaving only the innermost planet (shown in yellow) bound until the end of the integration.
Therefore, although the initial configuration satisfies the Hill stability criterion \citep{1993Icar..106..247G}, in the long term, planet-planet interactions can still lead to mild instabilities. When the same configuration evolves under the influence of a binary companion, secular perturbations amplify these interactions, producing outcomes consistent with the observed systems.

Once reduced to a single planet, the absence of apsidal precession allows capture into the vZLK resonance \citep{Takeda_2008,2011A&A...533A...7B}: $\omega_1$ transitions to a slow libration around a fixed value. This produces large, coupled oscillations in eccentricity and inclination. Prior to the ejections, mutual interactions forced rapid periapse precession, suppressing the resonance (all $\omega_j$ circulate), keeping the eccentricity low and suppressing the vZKL oscillations. Yet these same interactions gradually increased the eccentricity of the surviving planet, enabling Kozai cycles to emerge, $\omega_1$ then begins to librate around $\pi/2$. Following the removal of the outer planets, the eccentricity of the inner planet undergoes a sudden jump, nearly reaching unity, and subsequently oscillates between high and low values—anti-correlated with inclination, as expected from the vZLK mechanism, and $i_{p}$ oscillates between $i_{\rm KL}\simeq39^\circ$ and $i_{\rm B}$.

During these oscillations, the semi-major axis of the surviving planet remains nearly constant, while its eccentricity variations induce strong fluctuations in $q$ and $Q$. At peak eccentricities, the pericenter approaches very close to the host star, where short-range forces such as tides and general relativity \citep[GR,][]{2013ApJ...773..187N} drive additional periapse precession. When this precession counteracts that of the vZLK resonance, it can suppress further eccentricity growth. Nonetheless, in all simulations the planet's semi-major axis remains too large to trigger high-eccentricity tidal migration, preventing the formation of a warm or hot Jupiter, and also avoiding  tidal disruption \citep{2011ApJ...732...74G,2015ApJ...799...27P,2015ApJ...805...75P}.

The planetary orbits in our simulations never reach a maximum eccentricity sufficient to drive migration to a final semi-major axis where vZLK oscillations would be quenched. As a result, the cold Jupiters remain dynamically coupled to their stellar perturber, with their inclinations continuing to flip while maintaining approximately constant semi-major axes--in other words, the planets do not migrate. This outcome is consistent with theoretical expectations: according to \cite{Wu_2007}, the minimum inclination required to produce hot Jupiters is $81^\circ$, independent of $a_{\rm B}$. For sufficiently large inclinations, $q$ can reach very small values, allowing tidal dissipation to erode the planet's orbit--an outcome not obtained in our simulations.

The right panel in Fig.~\ref{fig:resonance_comparison} presents the phase-space portrait of the vZLK resonance in the representative $(\omega_1, e_1)$ plane, color-coded by the time elapsed since the system was reduced to a single planet. The system exhibits libration of the angular difference $\varpi_B - \varpi_1$ around $\pi$, a hallmark of secular interactions that indicates a coherent coupling between the orientations of the planetary and binary orbits \citep{2003ApJ...592.1201L}. In such configuration, the planet remains on a long-term stable orbit despite its high eccentricity and inclination, displaying dynamical behaviors rarely seen in isolated planetary systems.

Superimposed on the simulation is the octupole-level Hamiltonian \citep{Ford_2000,2011ApJ...742...94L}, valid for highly inclined and eccentric orbits, and is particularly well suited for comparison with our results. Initially, the innermost planet lies outside the libration region (solid lines); however, the instantaneous kick caused by the escape of the two outer planets pushes the surviving planet towards the stable fixed point (dashed lines in Fig.~\ref{fig:resonance_comparison}). There, it oscillates with large eccentricity amplitudes while maintaining stability. The pink line marks the separatrix, lying close to the integration points. The surviving planet sits at the boundary of the libration regime, and will remain captured in the resonant state in the absence of further perturbations. However, at such wide binary separations, galactic interactions may become significant, potentially destabilizing the system or causing the planet to migrate inward.

This behavior illustrates how the vZLK mechanism--absent in isolated star-planet systems--can dramatically reshape planetary orbits over long timescales, leading to high-eccentricity configurations with important implications for orbital stability and evolution of cold Jupiters.

To conclude, we quantify the outcomes of our simulations. Out of a total of 2500 simulations of a binary system initially hosting three planets on S-type orbits, the most common outcome was the survival of two or more planets, occurring in 1689 cases. This was followed by the survival of a single planet in 576 simulations. The least common result was the complete loss of all planets, with only 235 simulations ending with no surviving planets. These results highlight the relative efficiency of different dynamical pathways and show a clear tendency for systems to retain at least one planet despite the perturbations from a stellar companion.

The orbital evolution of the surviving planets was mainly shaped by a combination of planet-planet scattering and vZLK resonances, whose relative importance depended on the binary parameters. In close binaries ($a_{\rm B} \lesssim 100$\,au), scattering dominated the dynamics across most eccentricities, typically without exciting extreme planetary eccentricities. Wider binaries ($a_{\rm B} \gtrsim 200$\,au) only produced strong effects when highly eccentric ($e_{\rm B} \geq 0.75$), in which case scattering and secular resonances acted together. Inclination played a crucial role: for $i_{\rm B} \leq 30^\circ$ scattering prevailed, while at higher inclinations ($i_{\rm B} \geq 45^\circ$) secular effects frequently accompanied scattering, driving the surviving planets to higher eccentricities. In the most inclined cases ($i_{\rm B} \gtrsim 75^\circ$), vZLK resonances or temporary captures predominate, maintaining extremely eccentric orbits. Overall, our simulations show that high eccentricities and long-lived secular effects arise primarily in systems with inclined and eccentric binaries, whereas closer and more circular binaries are largely governed by scattering alone.

\subsection{Application to known exoplanetary systems}\label{sect:aplication_observed_systemas}

Fig.~\ref{fig:catalogo_estabilidad} presents 14 planetary systems hosting planets on highly eccentric orbits ($e_p \geq 0.8$). While our simulations--involving three Jupiter-mass planets orbiting a solar-mass star with a $0.8\,M_\odot$ stellar companion--aim to reproduce such configurations, not all observed systems share the physical characteristics or dynamical conditions assumed in our model, limiting the extent to which they can be matched.

Among the single-planet systems (represented by circles in Fig.\ref{fig:catalogo_estabilidad}), a few fall outside the regime our simulations can replicate. For instance, the estimated mass of the planet in \texttt{HD 26161~Ab} is $m_p = 28.5 \, M_{\rm Jup}$. This mass exceeds the commonly accepted threshold ($11-16 \, M_{\rm Jup}$) for deuterium burning, placing the object in the brown dwarf regime \citep{1997ApJ...491..856B,Spiegel_2011}. Thus, \texttt{HD 26161~Ab} likely qualifies as a brown dwarf, and the system should be interpreted as a hierarchical triple rather than a star-planet system, making it inconsistent with our simulated scenario.

A similar mass inconsistency arises with \texttt{HD 156846~Ab} and \texttt{HD 181234~Ab}, which have masses of $m_p \sim 10.5 \, M_{\rm Jup}$ and $m_p \sim 8.37 \, M_{\rm Jup}$, respectively. Although these values are below the brown dwarf threshold, they are still significantly larger than the planetary masses in our simulations. Since we started with three planets of $1 \, M_{\rm Jup}$, and even allowing for two mergers during evolution, the maximum mass achieved would only be $3 \, M_{\rm Jup}$. Reproducing a planet of $\sim 10.5 \, M_{\rm Jup}$ or $\sim 8.37 \, M_{\rm Jup}$ would require starting with much bigger planets, a scenario not explored in this study that might lead to different outcomes.

Given their extreme properties, four additional systems--\texttt{Gaia-3~Ab}, \texttt{TIC 241249530~b}, \texttt{HD 80606 b}, and \texttt{HR 5183 Ab}--cannot be fully accounted for by our current configurations. These systems contain planets with extremely high eccentricities ($e_p=0.84-0.95$) and distant stellar companions with semi-major axes between 1350 and 1667\,au. While some of these companions resemble our most distant scenarios ($a_{\rm B}=1250$ and $1500$\,au), in none of our tested configurations did they significantly perturb the planetary orbits. At such wide separations ($a_{\rm B}>1000$\,au), galactic tides are expected to play a dominant role.

The case of \texttt{HR 5183} is particularly striking: it hosts an extremely distant companion at $\sim15000$\,au, far beyond the range explored in our simulations. Nonetheless, \citet{2013Natur.493..381K} showed that even at such large distances, stellar encounters can still play a role in shaping its dynamical evolution.
Based on the observed orbital architecture, these planets should in principle fall within the GR-active regime and undergo migration driven by tidal effects. Yet, our integrations show otherwise--the planets remain essentially unperturbed. 
This discrepancy suggests that additional dynamical mechanisms must be considered, involving both processes acting in the inner regions near the host star and those associated with very distant companions.

\texttt{HD 80606}, in particular, has been extensively studied \citep[e.g.,][]{Wu_2003, Fabrycky_2007,2011CeMDA.111..105C}. With a periapsis distance of only 0.036\,au, the planet's orbit is still undergoing tidal evolution--a classic example of tidal-Kozai migration. The semi-major axis decreases through seemingly `discontinuous' steps because tidal dissipation becomes efficient only during episodes of extreme eccentricity, when the planet passes very close to the star. As the orbit contracts, the dominant source of periapsis precession gradually shifts from perturbations by the stellar companion to relativistic effects. This tidal evolution ultimately halts when the orbit fully circularizes, producing a typical hot Jupiter. Reaching such a configuration likely requires a mutual inclination of $\sim90^\circ$, a regime not covered by our current simulations.

\texttt{HD 144899~Ab} differs somewhat from the systems described above. The mass of the planet is consistent with those used in our simulations, as are the masses of both stars. However, the stellar companion has a semi-major axis of only 8\,au. Under the same initial conditions assumed for the planets, the companion would intrude into the system, most likely leading to a rapid instability. 

Finally, we have the cases of \texttt{GJ~3222} and \texttt{HD~4113}, both of which host two planets. In our simulations, systems that retained two planets after dynamical evolution typically exhibit similar, moderate eccentricities--unlike the pronounced eccentricity dichotomy seen in observed \texttt{GJ~3222} and \texttt{HD~4113}. In both systems, one planet has $e>0.9$, while its companion remains at $e\sim 0.5-0.6$. Moreover, \texttt{GJ 3222~Ac} and \texttt{HD 4113~Ac} have minimum masses above $50 \, M_{\rm Jup}$, placing them well within the brown dwarf regime and beyond the scope of our simulations. As a result, our model did not reproduce such extreme eccentricities in multiplanet systems, represented by star symbols in Fig.~\ref{fig:catalogo_estabilidad}. In fact, the highly eccentric inner planet in GJ 3222 is currently under Kozai migration, as its orbit lies in the GR-active zone where relativistic precession suppresses vZLK oscillations, driving a coupled decay of eccentricity and semi-major axis at nearly constant angular momentum.
\begin{figure}
\centering
    \includegraphics[width=\linewidth]{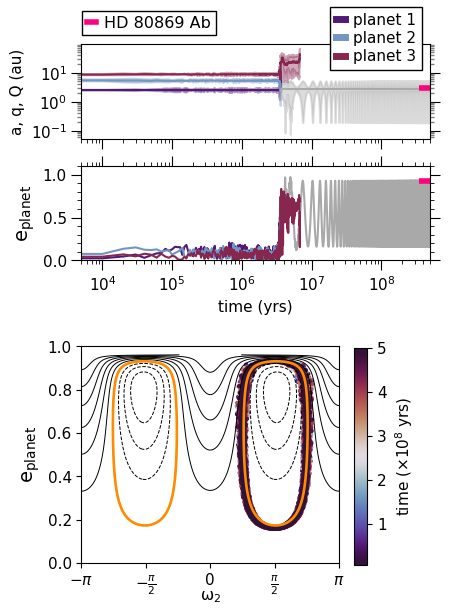}
    \caption{Temporal evolution of a \texttt{HD 80869}-like system, initialized with three Jupiter-mass planets. Adopted parameters for the stellar companion are $a_{\rm B} \simeq 250$\,au, $e_{\rm B} = 0.75$ and $i_{\rm B} = 45^{\circ}$, and a mass of $M_{\rm B} =0.6\, M_\odot$. From top to bottom we show the semi-major axis, eccentricity, and resonant plane for the system. The pink lines on the right indicate the observed semi-major axis and eccentricity of the planet.}
    \label{HD28254Ab}
\end{figure}

The remaining four of the 14 highly eccentric planets--\texttt{BD+63~1405~Ab}, \texttt{HD 80869~Ab}, \texttt{HD 28254~Ab}, and \texttt{HD 108341~Ab}--exhibit orbital characteristics consistent with our simulations. For each of these systems, we extracted the orbital elements from the Encyclopedia of Exoplanetary Systems and used them as a guide for selecting initial conditions. These systems serve as examples of how planet-planet scattering in the presence of a stellar companion, followed by the activation of the vZLK secular resonance, can generate highly eccentric single-planet configurations similar to those observed. Since the GR criterion depends on the binary mass, eccentricity, and inclination--of which only the mass is currently available for these systems--we evaluated it over the same ranges of eccentricity and inclination adopted in our simulations. In all cases, the planets lie outside the GR-active zone and, once the vZLK resonance is established, will continue undergoing oscillations in the absence of additional mechanisms.

As a proof of concept, we show the temporal evolution for the \texttt{HD 80869} system in Fig.~\ref{HD28254Ab}. \texttt{HD 80869} comprises a central star with mass $M_{\star} \approx 1.08\,M_{\odot}$ and a stellar companion of unknown mass, classified as an M-dwarf, with a projected semi-major axis of $a_{\rm B} = 249.61$\,au. The system hosts one known planet, \texttt{HD 80869~Ab}, with a semi-major axis of $a_p \approx  2.905$\,au, and an eccentricity of $e_p \approx 0.924$. These observed values are indicated by the pink lines in Fig.~\ref{HD28254Ab}. The planet's mass and orbital inclination remain unknown, but the minimum mass is estimated to be $m\,\mathrm{sin}(i) \approx 5 \,M_{\rm Jup}$.

Based on the results shown in Fig.~\ref{probabilidad_1}, a binary companion at $a_{\rm B} \simeq 250$\,au must have a high orbital eccentricity ($e_{\rm B} = 0.75-0.9$) to induce significant perturbations on the planets. To reproduce the \texttt{HD 80869} system, we performed a series of simulations assuming such high-eccentricity companions and tested various mutual inclinations ranging from coplanar to moderately inclined configurations ($i_{\rm B}=0-75^{\circ}$). Given the M-dwarf classification of the companion, we adopted a mass of $\rm{M_{\rm B}} =0.6 \, \rm{M_{\odot}}$, consistent with the expected range in $[0.075-0.6]\,\rm{M_{\odot}}$ for M-type stars  \citep{2000ApJ...542..464C}.

The system was initialized with three equal Jupiter-mass planets, adopting the same orbital setup as in our general simulations (see Section~\ref{sect:configuracion}). The observed orbital parameters of \texttt{HD 80869~Ab} were successfully reproduced in the run with $e_{\rm B}=0.75$ and $i_{\rm B}=45^{\circ}$. 
A first scattering event occurs at $\sim 2\times10^6$\,yr, producing a collision between the innermost and middle planets. A second instability takes place shortly before $10^7$\,yrs, during which the third planet is expelled from the system through its gravitational interaction with the surviving planet.
The final outcome is a single planet twice as massive (represented in gray) on a stable orbit characterized by $a_p \sim 2.8$\,au and $e_p^{\rm max}\sim 0.93$. 

The binary induces secular perturbations that lead to a permanent capture into the vZLK resonance (see bottom panel of Fig.~\ref{HD28254Ab}), which in turn drives the eccentricity to the observed values. It is worth noting that this system lies at the verge of the analytical stability limit given by the critical semi-major axis, $a_{\rm crit}$, for a single planet (see right panel of Fig.~\ref{fig:catalogo_estabilidad}), and it also resides close to the separatrix between libration and circulation (orange curve in the bottom panel of Fig.~\ref{HD28254Ab}). Nevertheless, our simulations show that the resonant configuration provides sufficient dynamical protection to stabilize the planet's orbit. At such large orbital distances, galactic tides are not expected to be efficient, and since both stellar tides and general relativistic precession are also ineffective in this regime, the planet should remain stable in its current state. Consequently, these planets are not expected to evolve into close-in configurations. 
As can be deduced from Fig.~\ref{probabilidad_1}, for the observed binary separation of $a_{\rm B} = 250$\,au, systems consistently end up with a single surviving planet whenever $e_{\rm B} \geq 0.75$. However, in order to reproduce the planet's eccentricity, it is also necessary that $i_{\rm B} > 30^{\circ}$.
 
It is worth mentioning that we were able to reproduce the remaining three systems with orbital features comparable to those observed (see Appendix \ref{more_systems}). To reproduce \texttt{BD+63 1405 Ab} we adopted $a_{\rm B}=97$\, au (as given by the observations), $e_{\rm B} = 0.5$ and $i_{\rm B} = 75^{\circ}$. Planet-planet scattering combined with temporary resonance capture, led to the survival of a planet with $a_p \sim 1.92$\,au and $e_p \sim0.825$. A similar mechanism operates in the simulation matching \texttt{HD 28254 Ab} ($a_{\rm B} = 240$\,au, $e_{\rm B} = 0.9$, and $i_{\rm B} = 15^{\circ}$), resulting in a planet with $a_p \sim 2.11$\,au and $e_p\sim 0.8$.
In contrast, the systems analogous to \texttt{HD~80869 Ab} ($a_{\rm B} = 250$\,au, $e_{\rm B} = 0.75$, and $i_{\rm B} = 45^{\circ}$) and \texttt{HD 108341 Ab} ($a_{\rm B} = 383$\,au, $e_{\rm B} = 0.9$, and $i_{\rm B} = 45^{\circ}$) experience eccentricity excitation via the vZLK mechanism, producing stable planets with $a_p \sim 2.8$\,au, $e_p \sim 0.93$ and $a_p \sim 2$\,au, $e_p \sim 0.85$, respectively--closely matching the observed values.

\section{Discussion}\label{sect:discussion}

The long-term stability of the highly eccentric planetary systems shown in Fig.~\ref{fig:catalogo_estabilidad} depends critically on the properties of the stellar companion. According to the analytical stability criterion derived by \citet{Holman_1999}, there exists a critical semi-major axis, $a_{\rm crit}$, beyond which planetary orbits become unstable due to the perturbations of the companion star. This threshold depends on the binary's eccentricity, the mass ratio, and the separation between the stars.

A major challenge in applying this criterion to observed systems lies in the limited knowledge of binary parameters. In most cases, only the projected separation of the companion is known, while key properties such as the binary's true eccentricity remain difficult to constrain--particularly for wide binaries with long orbital periods \citep{Hwang_2022}. Moreover, nearly half of wide binary systems host unresolved companions \citep{Tokovinin_2014, Moe_2017}, adding further uncertainty to the analysis. Although inclinations in wide binary systems have been measured and linked to the inclinations of their planets, the systems simulated in this work are not directly comparable to those observations. Most available data primarily concern close-in planets detected by transits, whereas our study focuses on planets at larger orbital distances, subject to different dynamical behaviors.

Fig.~\ref{probabilidad_1} highlights the critical role of the binary separation in determining the stability of planetary systems: close binaries tend to be highly disruptive, whereas wider binaries allow planets to survive over long timescales across a broader range of orbital configurations. While the analytical criterion suggests that planets with $a_p<a_{\rm crit}$ should remain stable, systems hosting multiple planets require full dynamical treatment. To this end, instead of relying on secular approximations, we follow the complete dynamical evolution of the system through direct N-body integrations. Our simulations show that planetary stability becomes strongly dependent on the binary's eccentricity and inclination, leading to dynamical instabilities even within the region that analytical estimates would deem stable.

These variations yield a diverse set of dynamical outcomes, ranging from highly unstable systems--where all planets are lost through ejection or collision--to fully stable systems in which all planets retain nearly unaltered orbits. To quantify these behaviors, we classify the outcomes of our 2500 simulations into three distinct categories: (i) zero surviving planets: the entire planetary system is destabilized either through ejection or collision, all planets are removed from the simulations (lower panels of Fig.~\ref{fig:prob_023}, representing 9\% of cases) (ii) one surviving planet: dynamical instabilities remove the rest of the system, leaving a single long-term survivor on a stable orbit (Fig.~\ref{probabilidad_1} and discussed in detail in Section~\ref{sect:general_results}, representing 23\% of the outcomes); and (iii) multiple surviving planets: two or more planets remain with only modest orbital perturbations (top and middle rows of Fig.~\ref{fig:prob_023}, representing 68\% of our simulations). This classification enables us to systematically assess how the binary parameters--particularly separation, eccentricity, and inclination--govern the long-term dynamical stability of planetary systems. 

Due to computational limitations, it is not feasible to explore the full range of companion orbital parameters through direct simulations. Instead, we fix the stellar masses to typical values and vary only the orbital parameters within the ranges $50 < a_{\rm B} < 1500$\,au, $0.05 < e_{\rm B} < 0.9$, and $0^{\circ} < i_{\rm B} < 75^{\circ}$. Future work could expand the parameter space to include a broader range of orbital configurations (such as those linked to the formation of hot Jupiters with $i_{\rm B}>81^\circ$), in addition to stellar and planetary masses. It is, however, to be expected that a lower mass stellar companions will have less influence, whereas more massive binaries will be dynamically more involved. Additional processes--such as tidal dissipation, general relativity, and galactic tides--may also influence the long-term dynamical evolution of planets in binaries, as well as their mutual interactions through mean-motion resonances \citep{2006ApJ...639..423M}, but will not modify the main conclusions of our work.

The binary configurations most likely to result in highly eccentric single-planet systems are predominantly associated with orbital inclinations above $45^\circ$, largely independent of the binary's semi-major axis. 
Our results are consistent with the mechanism proposed by \cite{2008ApJ...678..498N}, who studied the formation of close-in planets in isolated systems where planet-planet scattering can send one planet to a distant orbit, allowing it to act as a perturber. In their scenario, the eccentricity of the innermost planet can exceed a critical value, leading to strong tidal interactions with the central star and orbital circularization as the planet migrates inward. While we do not include stellar tides in our simulations, only about 1\% of our systems reach such extreme eccentricities, and therefore our conclusions would not be significantly altered by their inclusion.

Upcoming missions such as PLAnetary Transits and Oscillations of stars satellite \citep[PLATO][]{PLATO} and the Nancy Grace Roman Space Telescope \citep{Roman} will significantly improve our understanding of S-type planets by enabling new detections and providing the statistical basis to assess their prevalence. Complementary observations from missions like the CHaracterising ExOPlanets Satellite \citep[CHEOPS][]{CHEOPS} or JWST \citep{JWST} will further characterize their fundamental properties--such as radius, mass, and atmospheric composition--thereby refining current models of S-type planet formation and their dynamical evolution within binary systems. Follow-up observations of planet candidates, aimed at reducing contamination from false positives, together with measurements that constrain orbital architectures--such as studies of the Rossiter-McLaughlin effect to determine the projected obliquity--will provide critical insights into the relationship between planetary and orbital parameters. Furthermore, beyond detecting and characterizing planets in binary systems, forthcoming Gaia data releases should reveal additional binaries and refine their orbital properties, providing key observational constraints to inform and calibrate dynamical models of planetary systems in binaries.

It is important to note that, even if only a few systems are observed with high eccentricities, the coupling between eccentricity and inclination induced by the Kozai mechanism implies that we may be observing these systems during phases where the planets are on nearly circular but significantly inclined orbits. As a result, a larger population of systems may actually be undergoing Kozai-driven evolution, with their high-eccentricity phases simply missed due to observational biases or timing. This highlights the potential ubiquity of Kozai-like dynamics in shaping planetary architectures, even when not directly apparent in current observational samples.

\section{Conclusions} \label{sect:conclusiones}

Most stars form in binary or multiple systems, where planets must remain within a narrow range of stable orbits to withstand stellar perturbations. Numerous exoplanets have been discovered in such environments, making the study of their dynamics essential.

This paper emphasizes the pivotal role of gravitational interactions between stellar companions and their planets in shaping the long-term stability of these systems. Accurately capturing this complex dynamical behavior requires numerical simulations. Our work moves beyond simply assessing planetary stability, instead we have studied the particular sub-group of highly eccentric multiplanet systems to understand the role of a stellar binary in the orbital evolution of observed exoplanets in such systems. 

Our main findings can be summarized as follows:
\begin{enumerate}
    \item The companion star's orbital configuration critically shapes system stability, from preserving all planets with minimal orbital changes to causing complete system disruption, both in close and wide binaries.
    \item The binary eccentricity mainly determines how many planets survive, while the inclination controls the final eccentricities of the survivors.
    \item Binarity is essential for driving instabilities that excite planetary eccentricities beyond 0.8; planet-planet interactions alone cannot reproduce such extremes.
    \item Extremely high eccentricities occur only in systems with a single surviving planet.
    \item Multiple-planet systems emerge as one of the most common outcomes in our simulations, particularly for binaries with eccentricities $e_{\rm B} < 0.75$.
    \item Planet-planet scattering and secular processes--through the vZKL mechanism--often act together to produce abrupt orbital changes.
    \item Of the 14 observed systems, only 4 have parameters comparable to our simulations, and all of them were successfully reproduced.
    \item Despite reaching very high eccentricities and small periapses, the surviving planets' semi-major axes remain nearly constant, preventing migration into the Hot Jupiter region.
    \item We can predict that if more than one planet forms in a binary system, then as long as the stellar separation exceeds 200\,au, and the binary has an eccentricity $e_{\rm B} \leq 0.5$, a multi-planet system is expected to survive regardless of the binary's inclination.
\end{enumerate}

Future observations will likely provide more complete orbital information on stellar companions--beyond their projected separations--and reveal additional highly eccentric planets. Together, these advances could offer critical clues to reconstructing the dynamical histories of such systems.

\begin{acknowledgements}
The authors thank the anonymous reviewer for their  useful comments, which helped improve our manuscript. C.C. acknowledges support from Agencia Nacional de Investigación y Desarrollo (ANID) through FONDECYT post-doctoral grant n$^\circ$3230283. 
M.A.B and C.A.G. acknowledge support from FONDECYT Iniciación 11230741.
\end{acknowledgements}

\bibliographystyle{aa}
\bibliography{references}{}

\begin{appendix}
\section{Additional systems} \label{more_systems}

In Section \ref{sect:aplication_observed_systemas} we focused on \texttt{HD 80869} as a representative example. Here, we summarize the three additional systems with extreme eccentricities consistent with the parameters in our simulations.

\begin{figure}
    \centering
    \includegraphics[width=\columnwidth]{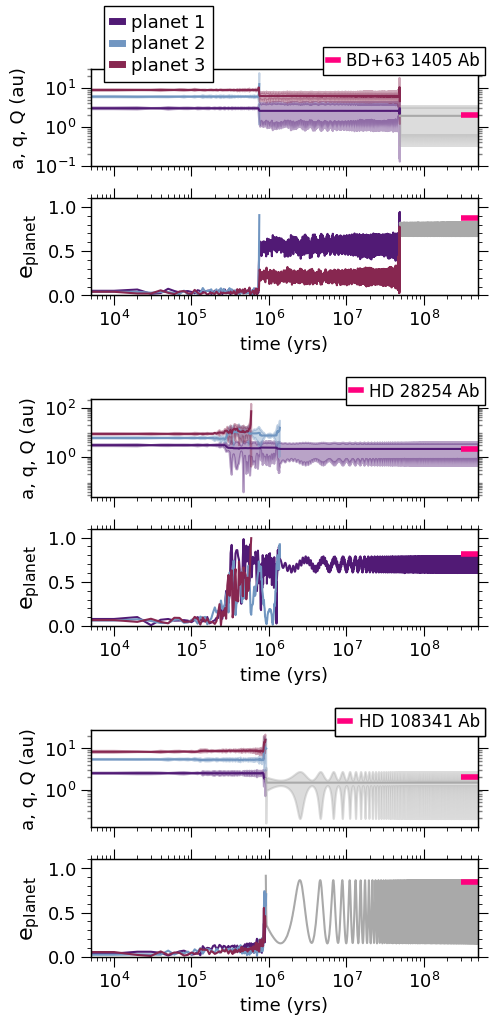}
    \caption{Temporal evolution of \texttt{BD+631405} (top), \texttt{HD 28254} (middle) and \texttt{HD 108341} (bottom).
    Upper panels show $a_{p}$, $q_{p}$, and $Q_{p}$, while lower panels show $e_{p}$. The pink lines on the right indicate the observed semi-major axis and eccentricity of the corresponding observed planet, BD+631405 Ab, HD 28254 Ab, and HD 108341 Ab.}
    \label{fig:otros_sistemas}
\end{figure}

The planet \texttt{BD+63 1405 Ab} has an eccentricity of $e_p\sim0.88$ and a semi-major axis of $a_p\sim 2.06$\,au, with a binary companion at $a_B\sim 97$\,au. According to Fig.\,\ref{fig:catalogo_estabilidad}, an extremely eccentric binary ($e_{\rm B}\sim0.9$) does not yield stable configurations at this separation. Fig.\,\ref{probabilidad_1} shows that for a binary at $a_{\rm B}\sim100$\,au, an eccentricity $0.05<e_{\rm B}<0.9$ is required to leave a single surviving planet. Additionally a highly inclined companion is needed to reach such eccentricities. Fig.\,\ref{fig:otros_sistemas} (top panels) shows the temporal evolution of a simulated system with $a_{\rm B}=97$ au, $e_{\rm B}=0.5$, and $i_{\rm B}=75^{\circ}$. In this case, the system evolves to a single planet with $e_{\rm p}\sim0.825$ and $a_{\rm p}\sim1.92$ au. The evolution is dominated by planet-planet scattering rather than vZKL resonances.

\texttt{HD 28254 Ab} has  $e_p\sim0.81$, and $a_p\sim2.15$ au, with a binary companion at $a_B\sim240$ au. Similarly, \texttt{HD 108341 Ab} has $e_p\sim0.85$, and $a_p\sim2$ au, with the binary at $a_B\sim383$ au. For binaries with $a_{\rm B}\sim250$–300 au, Fig.\,\ref{probabilidad_1} shows that $e_{\rm B}\geq0.75$ is required for the system to evolve toward a single surviving planet. Fig.\,\ref{fig:distribucion_inc_ecc} indicates that high binary inclination also favors extreme planetary eccentricities.

Fig.~\ref{fig:otros_sistemas} (middle panels) shows a simulated system similar to \texttt{HD 28254} with $a_{\rm B}=240$ au, $e_{\rm B}=0.9$, and $i_{\rm B}=15^{\circ}$. The system evolves to a single planet with $e_{\rm p}\sim0.8$ and $a_{\rm p}\sim2.11$ au, dominated by planet-planet scattering. The bottom panels show a \texttt{HD 108341}-like system with $a_{\rm B}=383$ au, $e_{\rm B}=0.9$, and $i_{\rm B}=45^{\circ}$, which evolves to a single planet with $e_{\rm p}\sim0.87$ and $a_{\rm p}\sim1.47$ au. In this case, the evolution is dominated by a combination of vZKL resonances and planetary interactions.

These four systems illustrate the critical role of binary eccentricity and inclination in producing extreme planetary eccentricities. While not all planets in Fig.~\ref{fig:catalogo_estabilidad} can be reproduced with our model, the consistency of these cases supports the conclusion that binary companions are a key driver of the observed extreme orbital configurations.

\section{Complementary results} \label{complementary}

\begin{figure*}
    \centering
    \includegraphics[width=\linewidth]{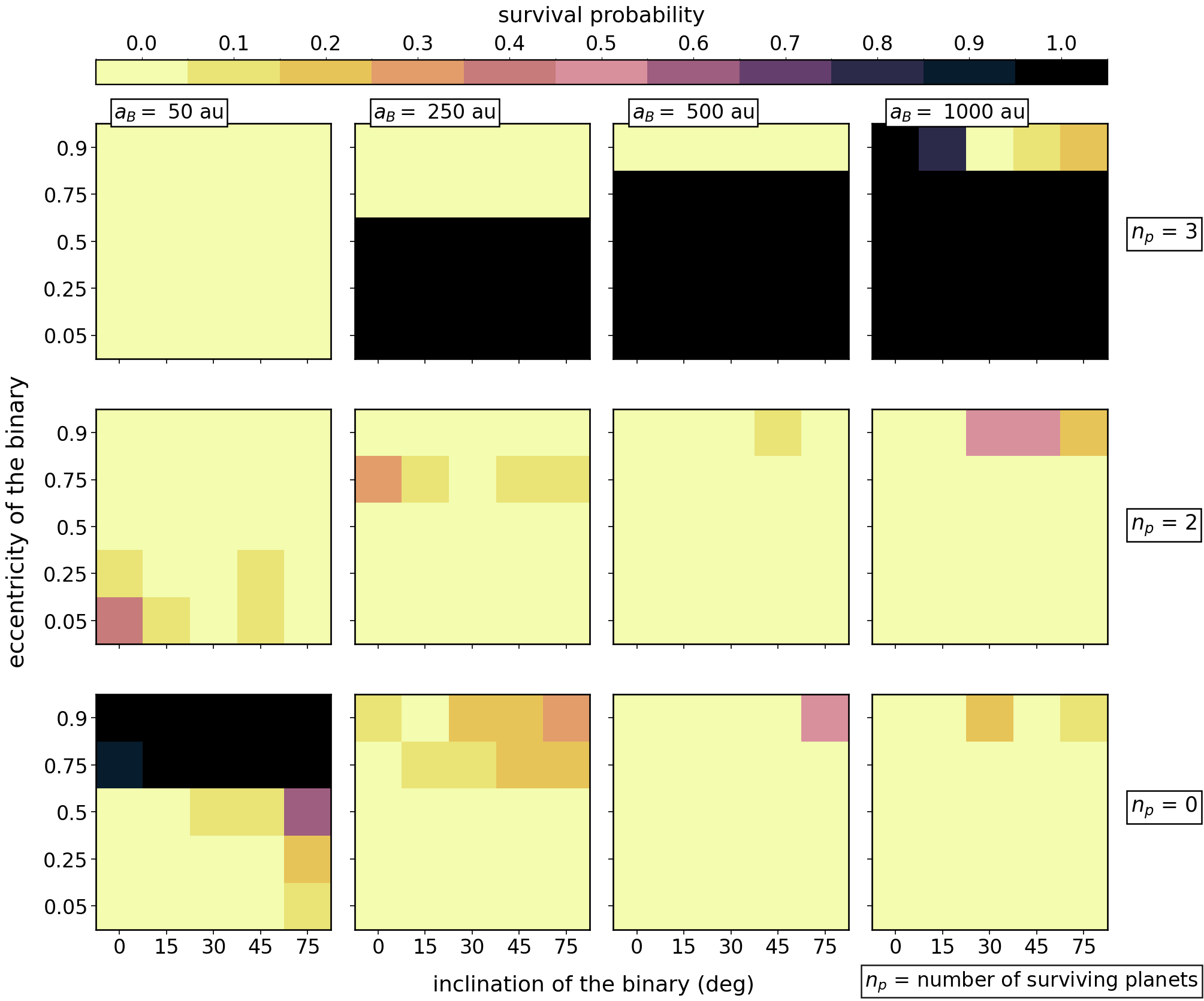}
    \caption{Survival probability distribution for three (top), two (middle), and zero planets (bottom). Only a subset of representative $a_{\rm B}$ values, where the behavior of the system changes significantly, are shown. This figure complements Fig.~\ref{probabilidad_1}, by illustrating the cases in which more than one planet survives or, conversely, the entire system becomes unstable.
    }
    \label{fig:prob_023}
\end{figure*}

To complement the parameter exploration presented in Section~\ref{sect:general_results}, which primarily focuses on the probability of obtaining a single surviving planet as a function of the binary's eccentricity and inclination, we present here a broader overview of the dynamical outcomes from the complete set of simulations. Specifically, we show the complementary cases for which 3, 2, or 0 planets survive (top, middle, and bottom panels of Fig.~\ref{fig:prob_023}, respectively), for representative values of the binary's semi-major axis. This more global perspective enables for a more comprehensive assessment of the system's stability under different combinations of the stellar companion's orbital parameters.

The top and bottom rows of Fig.~\ref{fig:prob_023} correspond to regimes identified as stable and unstable, respectively, and represent the same information shown in Fig.~\ref{probabilidad_1} through the diagonally and cross-hatched regions. These panels therefore offer a complementary visualization that highlights the combinations of parameters that systematically lead to long-term stability--or, conversely, to instability--of the system. The intermediate cases, in which one or two planets survive, illustrate the transition between these two extremes, with single-planet survival being more common than the survival of two.

\end{appendix}

\end{document}